\documentclass[12pt]{article}

\usepackage{epsf}
\title{Quantum cosmology with varying speed of light and Bohmian trajectories}
\author{F. Shojai, and S. Molladavoudi\\
Department of Physics, University of Tehran, Tehran, Iran.} 
\date{}
\begin{document}
\maketitle
\begin{abstract}
The classical trajectories for FLRW universe with varying speed of light are obtained for the cases in which the cosmological constant depends or not depend on the velocity of light. The theory is then quantized and the corresponding WDW equation is solved. It is shown that the method of causal interpretation of Bohm can be applied successfully to the theory. Finally the Bohmian trajectories are obtained and compared with the classical ones.
\end{abstract}
\section{Introduction}
In the early works of  Albrecht, 
Magueijo and Barrow \cite{alber,bar} for presenting a varying speed of light theory, the lorentz
invariance is broken and thus there is a preferred frame to
formulate the physical laws. These authors have 
assumed that there is a preferred frame in which the speed of
light is only a time dependent field and this frame is identified 
with the cosmological frame. Moreover in this theory a minimal coupling 
principle is assumed. According to it, in the preferred frame the 
curvature and the Einstein's tensor are computed by fixing 
the speed of light. Thus the Einstein's equations don't have any
correction terms in this frame and one should only replace a fixed
speed of light by a time--dependent one. Although this theory has the
ability to solve some of the problems of the standard Big--Bang cosmological 
models\cite{alber,mag}, it is not covariant and the comoving time is chosen 
as a specific time coordinate.

Therefore varying speed of light theory in its minimal formulation starts from the 
Einstein-Hilbert action in which the speed of light is substituted by a scalar field.
In addition the matter lagrangian is assumed to be explicitly
independent of the velocity of light and a term in the lagrangian is introduced which depends only on the velocity of light field in order to determine its dynamic.

In order to make more simplification, Barrow \cite{bar,barr} considers cases 
called pre--set $c$ in which the speed of light varies as a power of the 
expansion factor and so there is no dynamical term in the lagrangian for the velocity of light.

It is shown that it is possible to generalize these ideas \cite{mague} to have a varying speed of light theory, in addition to preserving the general covariance and local lorentz invariance. The price that have to be paid for this is to introduce a time--like coordinate $x^0$ which is not necessarily equal to $ct$. In terms of $x^0$ and $\vec{x}$, one has local lorentz invariance and general covariance. The physical time $t$, can only be defined when $dx^0/c$ is integrable.

Of course introduction of a varying speed of light in the Einstein-Hilbert action is not unique. This is because of the fact that multiplying it
by any power of light velocity and then introducing a varying speed of light  would give different 
results. This leads a more general theory\cite{mague}. Moreover the cosmological
constant generally depends on the velocity of light. If one assumes that it  
has field theoretical origin, it should scales as $c^4$. In general one 
can assume $c^n$ dependence for the cosmological constant \cite{barro}. (See \cite{magu} and references therein for a comprehensive review of varying speed of light theories.)

Using quantum cosmology with varying speed of light, the semiclassical tunneling probability in a minimal coupling model with
pre--set $c$ is disscused in \cite{har} and it is shown that it is different from the corresponding one in quantum 
cosmology with constant velocity of light. Furthermore in \cite{yu} the quantum cosmological
aspects of the minimal varying speed of light theory is discussed considering cosmological constant term and instantons.

On the other hand, applying Bohmian causal interpretation of quantum mechanics to quantum cosmology has several positive aspects compared to the copenhagen interpretation\cite{pin1,pin2,khod1}. For example, it provides time evolution of dynamical variables via Bohmian trajectories, whether the wavefunction depends or not on the time variable. It also leads to the possibility to avoid the initial singularity via the quantum force. It can lead to a graceful exit from inflation epoch. Investiagation of causal interpretation of quantum cosmological models with varying speed of light could be fruitful. It is interesting to see how the admission to the velocity of light to be a function of cosmological time can affect the Bohmian trajectories in early universe.

Here we shall consider the general action of \cite{mague,magu} and find it's exact classical
cosmological solution  for the flat universe and with or without cosmological constant, but without matter fields. After this we shall quantize this model and obtain the exact solutions of 
Wheeler--Dewitt equation for flat and non--flat and with or without cosmological constant. We shall show that it is possible to present a causal interpretation of the theory. The quantum trajectories are calculated and compared with the classical ones.
\section{The model}
The model for varying speed of light theory we use here is the one presented in \cite{mague}, with the action functional:
\begin{equation}
{\cal A}=\int d^4x \sqrt{-g}\left ( e^{\alpha\psi}\left ( {\cal R}-2\Lambda(\psi)-\gamma\nabla_\mu\psi\nabla^\mu\psi\right ) +e^{\beta\psi}{\cal L}_m(\phi_i,\partial_\mu\phi_i)\right )
\end{equation}
in which $e^\psi=c/c_0$, $c_0$ is a constant velocity and we put $8\pi G/c_0^4=1$. We also have a dynamical term for the velocity of light with the coupling constant $\gamma$, and $\phi_i$ represent matter fields. In the above action one assumes that the $c^4$ factor of Einstein--Hilbert action is broken into a part $c^\alpha$ behind the gravitational term of the action and a part $c^\beta$ behind the matter term. In order to remain faithful to the standard action of general relativity, one should set $\alpha-\beta=4$. Another point is that, here we have assumed that there is a time--like coordinate $x^0$, which is not equal to $ct$. In fact, since $c$ is a field, $cdt$ is not necessarily integrable. Therefore definition of physical time is only possible when one can integrate $dx^0/c$ \cite{mague}.

Here we shall deal with the minisuperspace of isotropic
and homogeneous cosmology. Denoting derivative with respect to the
coordinate $x^0$ with a dot, the action functional for a FLRW model would be:
\[
{\cal A}={\cal V}_3 \int dx^0 \left (e^{\alpha\psi} \left 
[ -6a\dot{a}^2-6\alpha a^2\dot{a}\dot{\psi}+12ka
-2\Lambda(\psi)a^3-\gamma a^3\dot{\psi}^2\right ]\right .
\]
\begin{equation}
\left . +2
a^3e^{\beta\psi}{\cal L}_m(\phi_i,\dot{\phi_i})\right )
\label{act}
\end{equation}
in which ${\cal V}_3$ is the three--space volume and $\Lambda(\psi)$, $a$ and $k$ are the cosmological constant, the scale factor and curvature parameter respectively. Depending on the nature of the cosmological constant one can distinguish two cases. First, gravitational cosmological constant $\Lambda_g$ which we assume that it is a fundamental constant of dimension $(\textit{length})^{-2}$ and hence it is $\psi$--independent. Second, field theoretical cosmological constant (i.e. one obtained from the field theoretical vacuum contribution to the cosmological constant) $\Lambda_f=\Lambda_0e^{n\psi}$. In this paper we shall choose $n=4$.

The corresponding canonical momenta are:
\begin{equation}
\Pi_a=\frac{\partial{\cal L}}{\partial \dot{a}}=-6a(2\dot{a}
+\alpha a\dot{\psi})e^{\alpha\psi}
\end{equation}
\begin{equation}
\Pi_\psi=\frac{\partial{\cal L}}{\partial \dot{\psi}}=-2a^2(
\gamma a\dot{\psi}+3\alpha\dot{a})e^{\alpha\psi}
\end{equation}
\begin{equation}
\Pi_{\phi_i}=\frac{\partial{\cal L}}{\partial \dot{\phi_i}}=
2 e^{\beta\psi}a^3\frac{\partial{\cal L}_m}{\partial 
\dot{\phi_i}}
\end{equation}
The Hamiltonian can be calculated as:
\[
{\cal H}=\frac{A_1e^{-\alpha\psi}}{a}\Pi_a^2
+\frac{A_2e^{-\alpha\psi}}{a^3}\Pi_\psi^2-
\frac{\alpha A_2e^{-\alpha\psi}}{a^2}\Pi_a\Pi_\psi
\]
\begin{equation}
-12kae^{\alpha\psi}+2\Lambda(\psi)a^3e^{\alpha\psi}+
2 e^{\beta\psi}a^3 {\cal T}
\label{ham}
\end{equation}
where:
\begin{equation}
A_1=\frac{-\gamma}{12}
\frac{2\gamma+3\alpha^2}{(2\gamma-3\alpha^2)^2}
\end{equation}
\begin{equation}
A_2=\frac{-3}{2}
\frac{2\gamma-\alpha^2}{(2\gamma-3\alpha^2)^2}
\end{equation}
\begin{equation}
{\cal T}(\phi_i,\dot{\phi_i})=\sum_i\left ( \dot{\phi_i}
\frac{\partial {\cal L}_m}{\partial\dot{\phi_i}}
-{\cal L}_m\right )
\end{equation}
where $\cal T$ is the zero--zero component of the energy-momentum tensor of the matter field.
\section{Classical trajectories}
Let us choose $k=0$ in this section. We shall distinguish three cases. By a $c$--dominated universe we mean ${\cal L}_m=\Lambda=0$. A $(c-\Lambda_g)$--dominated universe means that ${\cal L}_m=0$ but $\Lambda$ is not zero and is of gravitational kind, that is $\psi$--independent. Finally a $(c-\Lambda_f)$--dominated universe is one with ${\cal L}_m=0$ and $\Lambda=\Lambda_f(\psi)=\Lambda_0e^{4\psi}$. 
\subsection{$c$--dominated universe}
The classical equations of motion in this case can be derived from the action (\ref{act}) via putting ${\cal L}_m=0$ and $\Lambda=0$. They are:
\begin{equation}
2\frac{\ddot{a}}{a}+\frac{\dot{a}^2}{a^2}+\alpha\ddot{\psi}+2\alpha\frac{\dot{a}}{a}\dot{\psi}+(\alpha^2-\gamma/2)\dot{\psi}^2=0
\end{equation}
\begin{equation}
3\alpha\frac{\ddot{a}}{a}+3\alpha\frac{\dot{a}^2}{a^2}+\gamma\ddot{\psi}+3\gamma\frac{\dot{a}}{a}\dot{\psi}+\frac{\alpha\gamma}{2}\dot{\psi}^2=0
\end{equation}
These equations have the following solutions:
\begin{equation}
\frac{a}{a_0}=(x^0)^A
\end{equation}
\begin{equation}
\frac{c}{c_0}=e^\psi=(x^0)^B
\end{equation}
where $A$ and $B$ are one of these cases:
\begin{equation}
\left \{
\begin{array}{ll}
A=0 & B=0 \\
A=\frac{2(\alpha^2-\gamma)}{4\alpha^2-3\gamma} & B=\frac{2\alpha}{4\alpha^2-3\gamma} \\
A=\frac{\alpha^2 B-\alpha-\gamma B}{4\alpha^2B-\alpha-3\gamma B}& B=\frac{\alpha\pm\sqrt{9\alpha^2-6\gamma}}{4\alpha^2-3\gamma}\\
\end{array}
\right .
\end{equation}
The first solution is the flat Minkowski space--time, with constant speed of light. For the other two cases time can be defined as: 
\begin{equation}
t=\int\frac{dx^0}{c}=\frac{(x^0)^{1-B}}{c_0(1-B)}
\end{equation}
So that:
\begin{equation}
\frac{a}{a_0}=((1-B)c_0t)^{\frac{A}{1-B}}\sim t^{\frac{A}{1-B}}
\end{equation}
\begin{equation}
\frac{c}{c_0}=(\frac{a}{a_0})^{\frac{B}{A}}\sim t^{\frac{B}{1-B}}
\end{equation}
Depending on the values of $\alpha$ and $\gamma$ these can be either decreasing or increasing functions of time. In order to solve the horizon problem of cosmology, one should set \cite{magu} $\dot{a}>0$ and $\ddot{a}/\dot{a}-\dot{c}/c>0$. This leads to some limitations on $A$ and $B$ (and therefore on $\alpha$ and $\gamma$):
\begin{equation}
\frac{A}{1-B}<1, \ \ \ \ \ \ \frac{A-1}{1-B}>0
\end{equation}
\subsection{$(c-\Lambda_g)$--dominated universe}
The equations of motion are:
\begin{equation}
2\frac{\ddot{a}}{a}+\frac{\dot{a}^2}{a^2}+\alpha\ddot{\psi}+2\alpha\frac{\dot{a}}{a}\dot{\psi}+(\alpha^2-\gamma/2)\dot{\psi}^2-\Lambda_g=0
\end{equation}
\begin{equation}
3\alpha\frac{\ddot{a}}{a}+3\alpha\frac{\dot{a}^2}{a^2}+\gamma\ddot{\psi}+3\gamma\frac{\dot{a}}{a}\dot{\psi}+\frac{\alpha\gamma}{2}\dot{\psi}^2-\alpha\Lambda_g=0
\end{equation}
with the following solutions:
\begin{equation}
\frac{a}{a_0}=e^{px^0}
\end{equation}
\begin{equation}
\frac{c}{c_0}=e^\psi=e^{qx^0}
\end{equation}
where $p$ and $q$ are one of these four possibilities:
\begin{equation}
\left \{
\begin{array}{ll}
p=\frac{\alpha\sqrt{\Lambda_g}}{\sqrt{6\alpha^2-9\gamma/2}} & q=-\frac{\sqrt{6\Lambda_g}}{\sqrt{4\alpha^2-3\gamma}} \\
p=-\frac{\alpha\sqrt{\Lambda_g}}{\sqrt{6\alpha^2-9\gamma/2}} & q=\frac{\sqrt{6\Lambda_g}}{\sqrt{4\alpha^2-3\gamma}} \\
p=\frac{(\alpha^2-\gamma)\sqrt{\Lambda_g}}{\sqrt{6\alpha^4+3\gamma^2-17\gamma\alpha^2/2}} &
q=\frac{\alpha\sqrt{\Lambda_g}}{\sqrt{6\alpha^4+3\gamma^2-17\gamma\alpha^2/2}} \\
p=-\frac{(\alpha^2-\gamma)\sqrt{\Lambda_g}}{\sqrt{6\alpha^4+3\gamma^2-17\gamma\alpha^2/2}} &
q=-\frac{\alpha\sqrt{\Lambda_g}}{\sqrt{6\alpha^4+3\gamma^2-17\gamma\alpha^2/2}} 
\end{array}
\right .
\end{equation}
Although all of these solutions are acceptable, but the third one leads to anti--de Sitter space--time for $\alpha=0$ and $\gamma\rightarrow 0$ and the fourth one is de Sitter space--time in this limit.

Time can be defined as: 
\begin{equation}
t=\int\frac{dx^0}{c}=\frac{-e^{-qx^0}}{qc_0}
\end{equation}
So that:
\begin{equation}
\frac{a}{a_0}=(-qc_0t)^{\frac{-p}{q}}\sim t^{\frac{-p}{q}}
\end{equation}
\begin{equation}
\frac{c}{c_0}=(\frac{a}{a_0})^{\frac{q}{p}}\sim\frac{1}{t}
\end{equation}
In this case the speed of light  decreases as time increases. Again the horizon problem can be solved provided:
\begin{equation}
\frac{p}{q}<0
\end{equation}
\subsection{$(c-\Lambda_f)$--dominated universe}
The equations of motion are:
\begin{equation}
2\frac{\ddot{a}}{a}+\frac{\dot{a}^2}{a^2}+\alpha\ddot{\psi}+2\alpha\frac{\dot{a}}{a}\dot{\psi}+(\alpha^2-\gamma/2)\dot{\psi}^2-\Lambda_0e^{4\psi}=0
\end{equation}
\begin{equation}
3\alpha\frac{\ddot{a}}{a}+3\alpha\frac{\dot{a}^2}{a^2}+\gamma\ddot{\psi}+3\gamma\frac{\dot{a}}{a}\dot{\psi}+\frac{\alpha\gamma}{2}\dot{\psi}^2-\alpha\Lambda_0e^{4\psi}-4\Lambda_0e^{4\psi}=0
\end{equation}
with the solution:
\begin{equation}
\frac{a}{a_0}=(x^0)^s
\end{equation}
\begin{equation}
\frac{c}{c_0}=e^\psi=\frac{1}{\sqrt{x^0}}
\end{equation}
where $a_0$ and $s$ satisfy these algebraic equations:
\begin{equation}
3s^2-(\alpha+2)s+\frac{\alpha^2-\gamma/2}{4}-\Lambda_0a_0^{-2/s}=0
\end{equation}
\begin{equation}
6\alpha s^2-3(\alpha+\gamma/2)s+ \frac{\gamma(\alpha+4)}{8} - (\alpha+4) \Lambda_0a_0^{-2/s}=0
\end{equation}

In this case time is defined as: 
\begin{equation}
t=\int\frac{dx^0}{c}=\frac{2(x^0)^{3/2}}{3c_0}
\end{equation}
and thus:
\begin{equation}
\frac{a}{a_0}=(\frac{3}{2}c_0t)^{\frac{2s}{3}}\sim t^{\frac{2s}{3}}
\end{equation}
\begin{equation}
\frac{c}{c_0}=(\frac{a}{a_0})^{\frac{-1}{2s}}\sim t^{-\frac{1}{3}}
\end{equation}
Again the speed of light is a decreasing function of time, and the solution of horizon problem leads to:
\begin{equation}
s>1
\end{equation}
\section{Quantum solutions}
The classical model of previous sections can be quantized through the method of canonical quantization, i.e. by setting 
$\Pi_a\rightarrow -i\frac{\delta}{\delta a}$, 
$\Pi_\psi\rightarrow -i\frac{\delta}{\delta \psi}$ and
$\Pi_{\phi_i}\rightarrow -i\frac{\delta}{\delta \phi_i}$ in the hamiltonian (\ref{ham}). Here we set $\hbar=1$ and ignore factor ordering ambiguity. Since the theory is invariant under $x^0$ reparametrization, the Hamiltonian should be set equal to zero, and the 
WDW equation is then:
\[
-A_1\frac{\partial^2\Phi}{\partial a^2}
-\frac{A_2}{a^2}\frac{\partial^2\Phi}{\partial \psi^2}
+\frac{\alpha A_2}{a}
\frac{\partial^2\Phi}{\partial a\partial \psi}-12ka^2e^{2\alpha\psi}\Phi
\]
\begin{equation}
+2a^4\Lambda(\psi)e^{2\alpha\psi}\Phi+2 a^4 e^{(\alpha+\beta)\psi}
{\cal T}(\phi_i,-i\frac{\partial}{\partial \phi_i})\Phi=0
\label{wdw}
\end{equation}
in which $\Phi(a,\psi,\phi_i)$ is the wavefunctional.

Let us now try to solve the WDW equation for some cases.
\subsection{$c$--dominated solution}
The WDW equation would be:
\begin{equation}
A_1\frac{\partial^2\Phi}{\partial a^2}+\frac{A_2}{a^2}\frac{\partial^2\Phi}{\partial \psi^2}-\frac{\alpha A_2}{a}\frac{\partial^2\Phi}{\partial a\partial\psi}+12ka^2e^{2\alpha\psi}\Phi=0
\end{equation}
This equation can be solved via separation of variables method leading to the general solution:
\begin{itemize}
\item $k=0$
\begin{equation}
\Phi(a,\psi)=e^{i\omega\psi}a^{s}
\end{equation}
where 
\begin{equation}
A_2\omega^2+iA_2\omega\alpha s-A_1 s(s -1)=0
\end{equation}
The general solution is the linear combination of this solution.
\item $k\ne 0$
\begin{equation}
\Phi(a,\psi)=a^{c_1}J_\nu(c_2a^2e^{\alpha\psi})\ \ \ or\ \ \ a^{c_1}Y_\nu(c_2a^2e^{\alpha\psi})
\end{equation}
where 
\begin{equation}
c_1=\frac{2A_1}{4A_1-\alpha^2A_2}
\end{equation}
\begin{equation}
c_2^2=-72k
\end{equation}
\begin{equation}
\nu^2=\frac{2A_1^2(2A_1-\alpha^2A_2}{(4A_1-\alpha^2A_2)^3}
\end{equation}
\end{itemize}
\subsection{$(c-\Lambda_g)$--dominated solution}
Here we choose $k=0$. The WDW equation would be:
\begin{equation}
A_1\frac{\partial^2\Phi}{\partial a^2}+\frac{A_2}{a^2}\frac{\partial^2\Phi}{\partial \psi^2}-\frac{\alpha A_2}{a}\frac{\partial^2\Phi}{\partial a\partial\psi}-2\Lambda_g a^4e^{2\alpha\psi}\Phi=0
\end{equation}
with the solution:
\begin{equation}
\Phi=a^{c_1}J_\nu(c_2a^3e^{\alpha\psi})\ \ \ \textit{or}\ \ \ 
a^{c_1}Y_\nu(c_2a^3e^{\alpha\psi})
\end{equation}
where:
\begin{equation}
c_1=\frac{-3A_1}{\alpha^2A_2-6A_1}
\end{equation}
\begin{equation}
c_2^2= \frac{2\Lambda_g}{2\alpha^2A_2-9A_1}
\end{equation}
\begin{equation}
\nu^2=\frac{-c_1(c_1-1)A_1}{2\alpha^2A_2-9A_1}
\end{equation}
\subsection{$(c-\Lambda_f)$--dominated solution}
Again choosing $k=0$. The WDW equation would be:
\begin{equation}
A_1\frac{\partial^2\Phi}{\partial a^2}+\frac{A_2}{a^2}\frac{\partial^2\Phi}{\partial \psi^2}-\frac{\alpha A_2}{a}\frac{\partial^2\Phi}{\partial a\partial\psi}-2\Lambda_0 a^4e^{(4+2\alpha)\psi}\Phi=0
\end{equation}
with the solution:
\begin{equation}
\Phi=a^{c_1}J_\nu(c_2a^3e^{(2+\alpha)\psi})\ \ \ \textit{or}\ \ \ 
a^{c_1}Y_\nu(c_2a^3e^{(2+\alpha)\psi})
\end{equation}
where:
\begin{equation}
c_1=\frac{3A_1-2\alpha}{6A_1-2\alpha A_2-\alpha^2A_2}
\end{equation}
\begin{equation}
c_2^2= \frac{2\Lambda_0}{-9A_1+2(\alpha-1)(\alpha+2)A_2}
\end{equation}
\begin{equation}
\nu^2=\frac{-c_1(c_1-1)A_1}{-9A_1+2(\alpha-1)(\alpha+2)A_2}
\end{equation}
\section{The method of causal interpretation}
It is shown by de-Broglie and Bohm\cite{bohm} that it is possible to build a causal interpretation on the top of quantum theory. This is achieved by adding some axioms to quantum theory and removing some others. In de-Broglie--Bohm theory, the state of the system is denoted by $(\Phi(q),q(t))$ in which $q$ represents the degrees of freedom of the system, $\Phi$ is the wavefunction, and $q(t)$ is the trajectory of the system. The wavefunction dynamics is governed by Schr\"odinger equation, while the trajectory is defined via the \textit{guidance relation}, $\partial S/\partial q=\Pi$. $\Pi$ is the conjugate momentum of $q$, and $S$ is $\hbar$ times the phase of the wavefunction. Therefore what is added to quantum theory is that the state of the system is not determined by the wavefunction only, but one should add the trajectory of the system to it. 

On the other hand in de-Broglie--Bohm theory one does not need to use the projection axiom of the quantum theory. In fact, it can be shown easily\cite{bohm,measure} that when some property of the system is measured, the trajectory of the system is such that after a very short time the state of the system is one of eigenstates of the observable and the probability of finding any eigenvalue is given by the Born law. Therefore de-Broglie--Bohm theory has not the measurement axiom.

In this way, the quantum theory of de-Broglie--Bohm, which we shall call causal quantum theory, is really a different theory from quantum theory. 

All of these things are motivated by decomposing the Schr\"odinger equation into two parts, by writing $\Phi=R \exp (iS/\hbar)$. In this way one gets two equations. One of them is the continuity equation with $\rho=R^2$ as the ensemble density and the other one is a Hamilton--Jacobi equation, with some additional term called \textit{quantum potential} which is responsible for quantum behaviours.

For our model, the causal interpretation can be derived very easily via putting
\begin{equation}
\Phi(a,\psi)=R(a,\psi)\exp (iS(a,\psi)/\hbar)
\end{equation}
in equation (\ref{wdw}). Here we neglected matter part for simplicity. The result is:
\begin{equation}
A_1\frac{\partial}{\partial a}(R^2\frac{\partial S}{\partial a})+
A_2\frac{\partial}{a^2\partial\psi}(R^2\frac{\partial S}{\partial\psi})-
\alpha A_2\frac{\partial}{2a\partial\psi}(R^2\frac{\partial S}{\partial a})-
\alpha A_2\frac{\partial}{2a\partial a}(R^2\frac{\partial S}{\partial\psi})=0
\end{equation}
\begin{equation}
{\cal H}(\partial S/\partial a, \partial S/\partial \psi; a, \psi)+{\cal Q}=0
\end{equation}
where the first term of the latter equation is the Hamiltonian (\ref{ham}) in which $\Pi_a$ and $\Pi_\psi$ are replaced by $\partial S/\partial a$ and  $\partial S/\partial \psi$ respectively. This is the Hamilton--Jacobi equation with the additional quantum potential term:
\begin{equation}
{\cal Q}=\frac{-\hbar^2e^{-\alpha\psi}}{aR}\left ( A_1\frac{\partial^2R}{\partial a^2}+\frac{A_2}{a^2}\frac{\partial^2R}{\partial\psi^2} -\frac{\alpha A_2}{a}\frac{\partial^2R}{\partial a\partial\psi}\right )
\end{equation}
The Bohmian trajectories can be evaluated via the guidance relations:
\begin{equation}
\Pi_a=\frac{\partial{\cal L}}{\partial \dot{a}}=-6a(2\dot{a}
+\alpha a\dot{\psi})e^{\alpha\psi}=\frac{\partial S}{\partial a}
\end{equation}
\begin{equation}
\Pi_\psi=\frac{\partial{\cal L}}{\partial \dot{\psi}}=-2a^2(
\gamma a\dot{\psi}+3\alpha\dot{a})e^{\alpha\psi}=\frac{\partial S}{\partial\psi}
\end{equation}
\section{Bohmian trajectories}
Bohmian trajectories highly depends on the wavefunction of the system. For example if the wavefunction is real, the guidance relation leads to zero conjugate momenta. But if one chooses some general linear combination of real solutions, one would get a non--trivial trajectory. This shows that one cannot speak of Bohmian trajectories and their difference with the classical ones without specifying that which linear combination of independent solutions of the WDW equation is chosen to be the wavefunction. One criteria for choosing the wavefunction could be having classical limit. That is one may fix the linear combination via requiring that for large times, say, the Bohmian trajectories tends to the classical one, in other words the quantum potential is ignorable in comparison to the classical potential for large times\cite{measure,limit}. It must be noted that in general, it is not possible to find solutions with classical limit, i.e. making the quantum potential going to zero. As an example, it is shown recently\cite{ooo} that for nonseparable dynamical quantum systems Bohmian trajectories never can approach the classical ones.

In what follows we shall choose specific wavefunctions and investigate the corresponding Bohmian trajectories for flat universe to demonstrate how these trajectories can be. 
\subsection{$c$--dominated universe}
Let us choose the wavefunction of this case as:
\begin{equation}
\Phi=a^se^{i\omega\psi}
\end{equation}
which leads to the conjugate momenta:
\begin{equation}
\Pi_a=0
\end{equation}
\begin{equation}
\Pi_\psi=\omega
\end{equation}
and thus the Bohmian trajectories are given by equations:
\begin{equation}
\dot{a}=\frac{\alpha\omega}{2(2\gamma-3\alpha^2)}\frac{e^{-\alpha\psi}}{a^2}
\end{equation}
\begin{equation}
\dot{\psi}=\frac{-\omega}{2\gamma-3\alpha^2}\frac{e^{-\alpha\psi}}{a^3}
\end{equation}
These two equations can be combined to give the relations
\begin{equation}
\frac{a}{a_0}=e^{-\alpha\psi/2}
\end{equation}
where $a_0$ is a constant of integration and 
\begin{equation}
\dot{a}=\frac{\alpha\omega}{2a_0^2(2\gamma-3\alpha^2)}=C=\textit{constant}
\end{equation}
Therefore we have 
\begin{equation}
a=a_0+Cx^0
\end{equation}
\begin{equation}
\frac{c}{c_0}=\left ( \frac{a_0}{a_0+Cx^0}\right )^{2/\alpha}
\end{equation}
Since $c$ is a function of $x^0$ only, the time can be defined as before:
\begin{equation}
t=\frac{1}{Cc_0a_0^{2/\alpha}(1+2/\alpha)}(a_0+Cx^0)^{1+2/\alpha}
\end{equation}
In terms of this physical time we have:
\begin{equation}
\frac{a}{a_0}\sim t^{1/(1+2/\alpha)}
\end{equation}
\begin{equation}
\frac{c}{c_0}\sim t^{-2/(2+\alpha)}
\end{equation}
Comparison with the classical solution shows that it could or could not be the classical solution. In fact there are specific choices of $\alpha$ and $\gamma$ that the Bohmian trajectories and the classical ones coincides. 
This solution is a contracting universe which has no horizon problem ($\dot{a}<0$ and $\ddot{a}/\dot{a}-\dot{c}/c<0$).

The quantum potential for this solution is:
\begin{equation}
{\cal Q}=-A_1s(s-1)a^{-3}e^{-\alpha\psi}\sim t^{-1/(1+2/\alpha)}
\end{equation}
\subsection{$(c-\Lambda_g)$--dominated universe}
Now let us choose a more realistic wavefunction for this case, a wavefunction which leads to trajectories tending to the classical ones for large scale factors. Using the asymptotic forms of Bessel functions:
\begin{equation}
J_\nu(x)\sim\sqrt{\frac{2}{\pi x}}\cos (x-\nu\pi/2-\pi/4)
\end{equation}
\begin{equation}
Y_\nu(x)\sim\sqrt{\frac{2}{\pi x}}\sin (x-\nu\pi/2-\pi/4)
\end{equation}
One observes that a linear combination of the form:
\begin{equation}
\Phi=a^{c_1}\left ( J_\nu(c_2a^3e^{\alpha\psi})+iY_\nu(c_2a^3e^{\alpha\psi})\right )
\label{asd}
\end{equation}
has correct limit. For large $a^3e^{\alpha\psi}$ the phase of the wavefunction is:
\begin{equation}
S=c_2a^3e^{\alpha\psi}+\textit{constant}
\end{equation}
so that:
\begin{equation}
\Pi_a=3c_2a^2e^{\alpha\psi}
\end{equation}
\begin{equation}
\Pi_\psi=\alpha c_2 a^3e^{\alpha\psi}
\end{equation}
and the equations of motion are:
\begin{equation}
\dot{a}=\frac{1}{2}c_2(\alpha^2-\gamma)a
\end{equation}
\begin{equation}
\dot{\psi}=\frac{1}{2}\alpha c_2
\end{equation}
with solutions:
\begin{equation}
\frac{a}{a_0}=e^{\frac{1}{2}c_2(\alpha^2-\gamma)x^0}
\end{equation}
\begin{equation}
\frac{c}{c_0}=e^{\frac{1}{2}\alpha c_2 x^0}
\end{equation}
Time is equal to:
\begin{equation}
c_0t=-\frac{2}{\alpha c_2}e^{-\frac{1}{2}\alpha c_2 x^0}
\end{equation}
So that in terms of time we have:
\begin{equation}
\frac{c}{c_0}=-\frac{2}{\alpha c_0 c_2t}\sim \frac{1}{t}
\end{equation}
\begin{equation}
\frac{a}{a_0}=\left (-\frac{1}{2}c_0 c_2 t\right )^{-\frac{\alpha^2-\gamma}{\alpha}}\sim t^{-\frac{\alpha^2-\gamma}{\alpha}}
\end{equation}
Comparison with the classical solution shows that we have the correct limit. The quantum potential is now:
\begin{equation}
{\cal Q}=(-A_1(c_1-3/2)(c_1-5/2)-\alpha^2 A_2(c_1-1)/2)a^{-3}e^{-\alpha\psi}
\end{equation}
which is ignorable in comparison to the classical potential when $a^3e^{\alpha\psi}$ is large.

For small $a^3e^{\alpha\psi}$, using the expansions:
\begin{equation}
J_\nu(x)=\frac{2^{-\nu}}{\Gamma(1+\nu)}x^\nu+\cdots
\end{equation}
\begin{equation}
Y_\nu(x)=-\frac{2^\nu}{\Gamma(1-\nu)}\frac{1}{\sin\pi\nu}x^{-\nu}+ \frac{2^{-\nu}}{\Gamma(1+\nu)}\cot\pi\nu x^\nu+\cdots
\end{equation}
The phase of the wavefunction is
\begin{equation}
\tan S=-\frac{2^{2\nu}\Gamma(1+\nu)}{\Gamma(1-\nu)\sin\pi\nu}a^{-6\nu} e^{-2\alpha\nu\psi}\equiv Ca^{-6\nu}e^{-2\alpha\nu\psi}
\end{equation}
The canonical momenta are then:
\begin{equation}
\Pi_a=-\frac{6\nu}{C}a^{6\nu-1}e^{2\alpha\nu\psi}
\end{equation}
\begin{equation}
\Pi_\psi=-\frac{2\alpha\nu}{C}a^{6\nu}e^{2\alpha\nu\psi}
\end{equation}
and the equations of motion are:
\begin{equation}
\dot{a}=Ba^{6\nu-2}e^{\alpha(2\nu-1)\psi}
\end{equation}
\begin{equation}
\dot{\psi}=B'a^{6\nu-3}e^{\alpha(2\nu-1)\psi}
\end{equation}
in which:
\begin{equation}
B=\frac{\nu}{C}\frac{\gamma-\alpha^2}{2\gamma-3\alpha^2}
\end{equation}
and
\begin{equation}
B'=-\frac{\alpha\nu}{C}\frac{1}{2\gamma-3\alpha^2}
\end{equation}
Dividing these two equations we have
\begin{equation}
\frac{a}{a_0}=e^{\frac{B}{B'}\psi}=\left (\frac{c}{c_0}\right )^{\frac{B}{B'}}
\end{equation}
And the solution is:
\begin{equation}
a\sim t^{\delta_1}
\end{equation}
\begin{equation}
\psi\sim t^{\delta_2}
\end{equation}
where
\begin{equation}
\delta_1=\frac{\gamma-\alpha^2}{(3\gamma+\alpha-4\alpha^2)+2\nu(4\alpha^2-3\gamma)}
\end{equation}
\begin{equation}
\delta_2=\frac{-\alpha}{(3\gamma+\alpha-4\alpha^2)+2\nu(4\alpha^2-3\gamma)}
\end{equation}
The quantum potential can be calculated as:
\begin{equation}
{\cal Q}=(-A_1(c_1-3\nu)(c_1-3\nu-1)-\alpha^2 \nu^2 A_2-\alpha^2 \nu A_2 (c_1-3\nu))a^{-\nu-2}e^{-\alpha\psi}
\end{equation}
In order to illustrate how could the Bohmian trajectories be, we present here some plots of trajectories and quantum potential. First we choose $\gamma=5$ and $\alpha=-2$. The quantum potential for this case is plotted in figure (\ref{fig1}).
\epsfxsize=5in
\epsfysize=3.98in
\begin{figure}[htb]
\begin{center}
\epsffile{bvslfig1.eps}
\end{center}
\caption{Plot of quantum potential as a function of the normalized scale factor ($a/a_0$) and normalized velocity of light ($c/c_0$), for a $(c-\Lambda_g)$--dominated universe. For this graph we set $\gamma=5$ and $\alpha=-2$.}
\label{fig1}
\end{figure}
The classical and quantum trajectories of the scale factor and the velocity of light are plotted in figures (\ref{fig2}) and (\ref{fig3}). It is clear that the trajectories tends to classical ones for large times. In this classical limit the quantum potential goes to zero. 
\epsfxsize=5in
\epsfysize=3.15in
\begin{figure}[htb]
\begin{center}
\epsffile{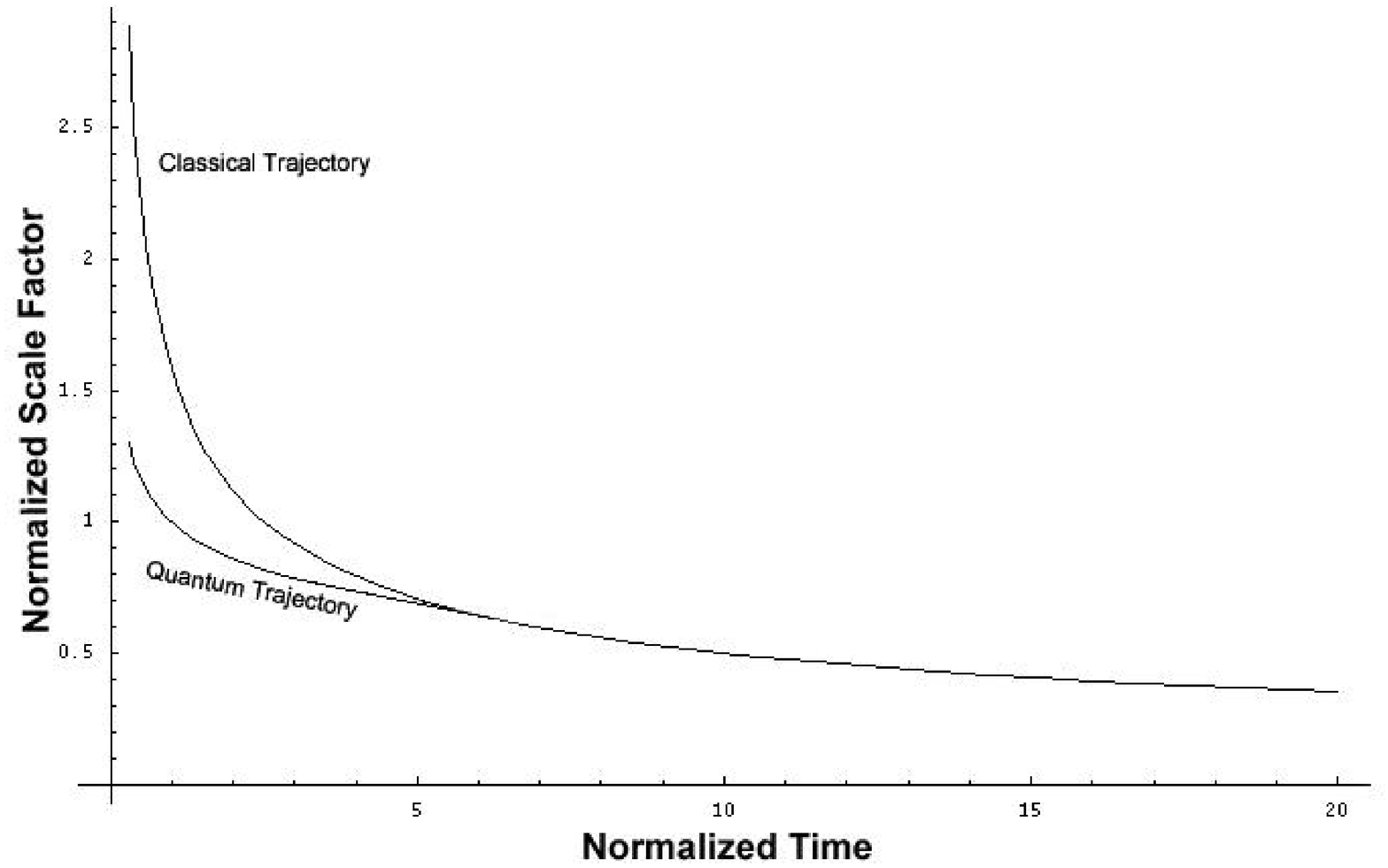}
\end{center}
\caption{Classical and quantum trajectories of the normalized scale factor ($a/a_0$), for a $(c-\Lambda_g)$--dominated universe. For this graph we set $\gamma=5$ and $\alpha=-2$. Time axis is normalized to $0.4\sqrt{\Lambda_g}c_0$.}
\label{fig2}
\end{figure}
\epsfxsize=5in
\epsfysize=3.2in
\begin{figure}[htb]
\begin{center}
\epsffile{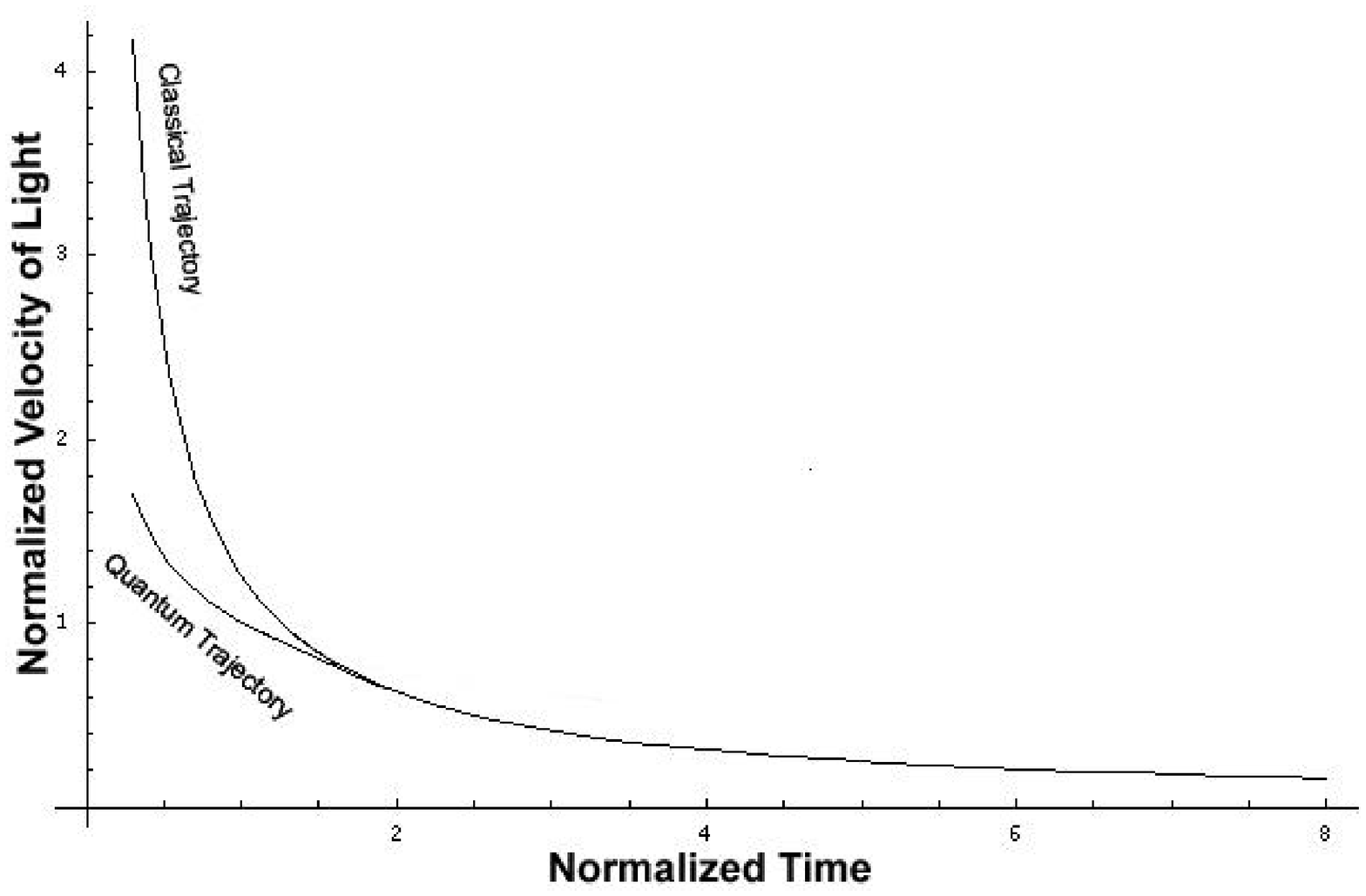}
\end{center}
\caption{Classical and quantum trajectories of the normalized velocity of light ($c/c_0$), for a $(c-\Lambda_g)$--dominated universe. For this graph we set $\gamma=5$ and $\alpha=-2$. Time axis is normalized to $0.4\sqrt{\Lambda_g}c_0$.}
\label{fig3}
\end{figure}

As a second example we choose $\gamma=2.5$ and $\alpha=1$. The quantum potential, the classical and quantum trajectories of the scale factor and the velocity of light are plotted in figures (\ref{fig4}), (\ref{fig5}) and (\ref{fig6}). Again the trajectories tends to classical ones for large times and  the quantum potential goes to zero.
\epsfxsize=5in
\epsfysize=3.95in
\begin{figure}[htb]
\begin{center}
\epsffile{bvslfig4.eps}
\end{center}
\caption{Plot of quantum potential as a function of the normalized scale factor ($a/a_0$) and normalized velocity of light ($c/c_0$), for a $(c-\Lambda_g)$--dominated universe. For this graph we set $\gamma=2.5$ and $\alpha=1$.}
\label{fig4}
\end{figure}
\epsfxsize=5in
\epsfysize=3.3in
\begin{figure}[htb]
\begin{center}
\epsffile{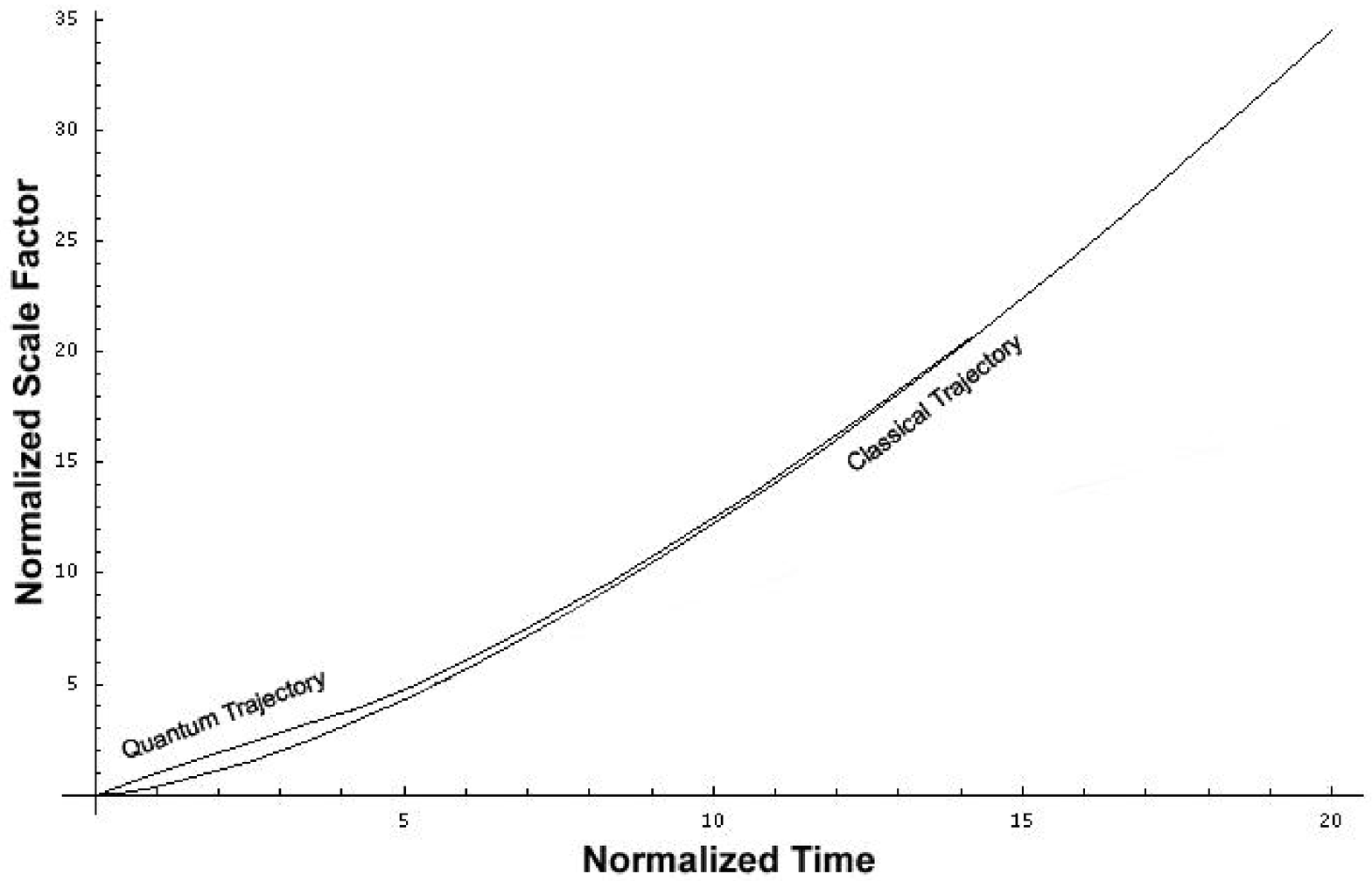}
\end{center}
\caption{Classical and quantum trajectories of the normalized scale factor ($a/a_0$), for a $(c-\Lambda_g)$--dominated universe. For this graph we set $\gamma=2.5$ and $\alpha=1$. Time axis is normalized to $0.5\sqrt{\Lambda_g}c_0$.}
\label{fig5}
\end{figure}
\epsfxsize=5in
\epsfysize=3.27in
\begin{figure}[htb]
\begin{center}
\epsffile{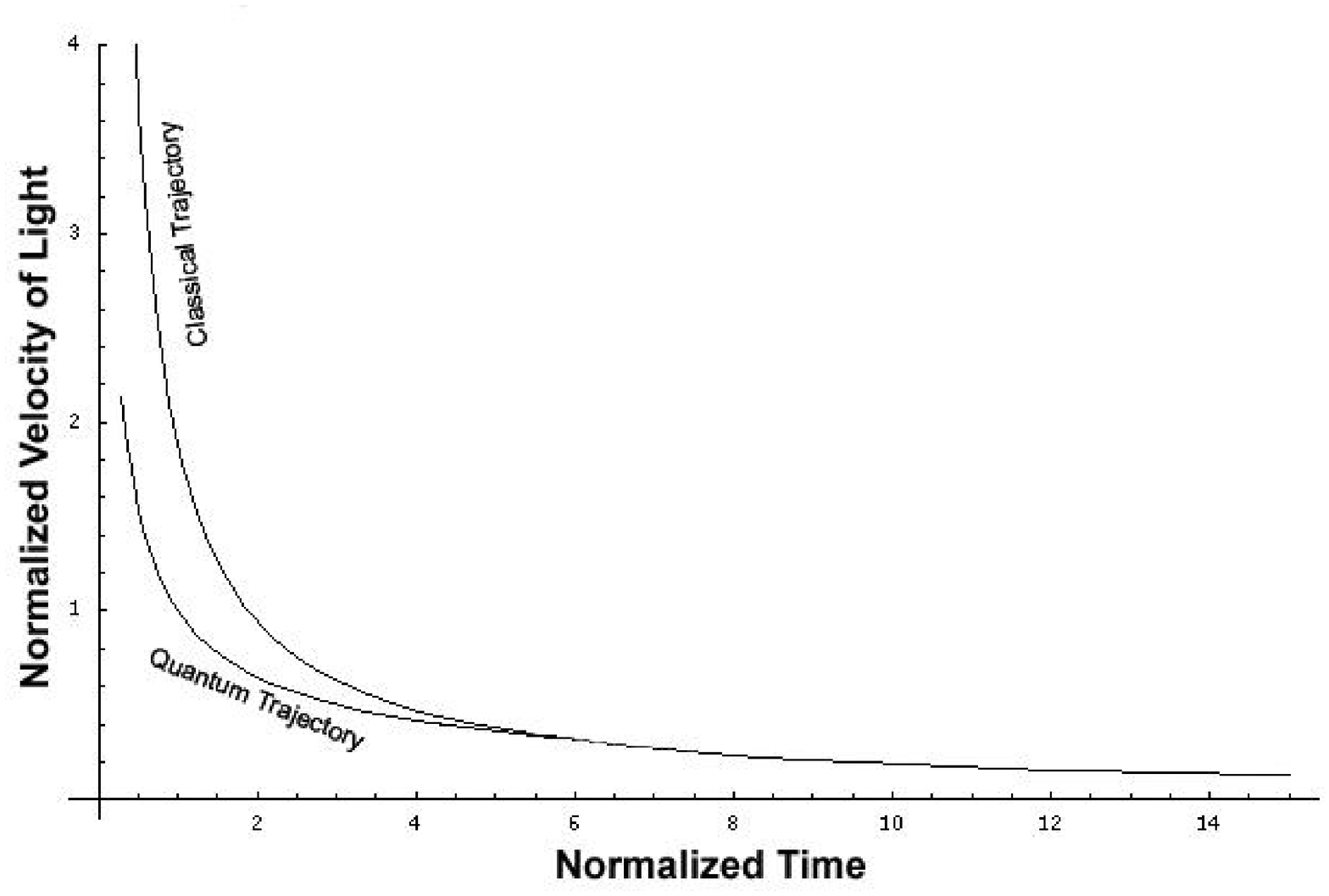}
\end{center}
\caption{Classical and quantum trajectories of the normalized velocity of light ($c/c_0$), for a $(c-\Lambda_g)$--dominated universe. For this graph we set $\gamma=2.5$ and $\alpha=1$. Time axis is normalized to $0.5\sqrt{\Lambda_g}c_0$.}
\label{fig6}
\end{figure}
\subsection{$(c-\Lambda_f)$--dominated universe}
Since the wavefunctions in this case are very similar to the previous case, so are the trajectories and we do not present it here.
\section{Conclusion and remarks}
In this paper we have obtained the Bohmian trajectories corresponding to a varying speed of light theory. This trajectories are shown that can have correct classical limits. We have done this for a matter free universe so that only we are able to see the status of the horizon problem for the trajectories. In order to disscuss about other cosmological problems like the flatness problem, one needs to extend the model to the universes with matter. For this we have to solve the WDW equation with matter fields and obtain the Bohmian trajectories using the method of \cite{aaa}. We shall do this in a forthcomming paper.

It must be noted that the specific choice of the linear combination in equation (\ref{asd}) imposes the boundary condition that the wavefunction behaves as an outgoing wave. This is what expected, because in a region where one has classical limit, one expects that Bohmian trajectories and the WKB path coincide. Therefore the appearance of an outgoing wave as the boundary condition is the sign of having the classical limit.

Another important thing should be noted here that although the investigation of whether these solutions of WDW equation have unitary evolution needs addition of matter fields (as it is discussed in the literature, see e.g. \cite{uni}), but since the the linear combination of the equation (\ref{asd}) have only outgoing WKB wave limit, it would have a unitary evolution, see \cite{uni}. For other linear combinations of the solutions it is suitable to choose a wavepacket with compact support to have unitary evolution.

We studied here the solutions which have the classical limit, but as the reader may noticed the classical singularity is not avoided e.g. in the figure (\ref{fig5}). It is possible to choose the parameters in such a way that the classical singularity is avoided but the classical limit is not present. The fact that quantum force has the ability to stop gravity from making things singular is wellknown and disscused in the literature\cite{pin3,khod2,pin4}. The possibility of having  nonsingular solutions with correct classical limit would be disscused in future works for more realistic cases when matter is present.

At this end, it should be noted that we have ignored factor ordering ambiguity. In the causal approach to quantum cosmology, although the form of quantum potential depends on the regularization and factor ordering, but there are some general results that are independent of the specific factor ordering chosen. For example it is shown\cite{pin5,aaa,pin6} that the quantum constraint algebra is independent of factor ordering. Also the general results obtained here do not change dramatically with respect to the factor ordering. This is the reason why we have ignored it in this paper.

\textbf{Acknowledgment} This work is partly supported by a grant from university of Tehran and partly by a grant from center of excellence of department of physics on the structure of matter.

\end{document}